\documentclass{article}

\usepackage[preprint]{neurips_2026}

\usepackage[utf8]{inputenc}
\usepackage[T1]{fontenc}
\usepackage{hyperref}
\usepackage{url}
\usepackage{booktabs}
\usepackage{amsmath}
\usepackage{amsfonts}
\usepackage{nicefrac}
\usepackage{microtype}
\usepackage{xcolor}
\usepackage{graphicx}
\usepackage{subcaption}
\usepackage{float}

\title{Agentic Hardware Design as Repository-Level \\Code Evolution}

\author{%
  Cunxi Yu\thanks{Corresponding author: \texttt{cunxiy@nvidia.com}} \\
  NVIDIA Research \\
  \And
  Chenhui Deng \\
  NVIDIA Research \\
  \And
  Nathaniel Pinckney \\
  NVIDIA Research \\
  \And
  Brucek Khailany \\
  NVIDIA Research \\
}

\newcommand{\ours}{HORIZON}

\begin{document}

\maketitle

\begin{abstract}
We present \ours{}, a self-evolving agent framework that treats hardware design as repository-level code evolution. A Markdown harness is compiled into a project pack containing domain knowledge, an executable evaluator, an acceptance predicate, and a git/runtime policy; a hands-free agent loop then evolves an isolated git worktree, using repository operations for state management, tracing, and replay. This extends prior works of repository-scale self-evolution from EDA software systems, to hardware-design artifacts themselves. We evaluate our approach on ChipBench, RTLLM, Verilog-Eval, and nine CVDP categories, achieving 100\% benchmark completion across all suites with a fully hands-free agentic loop. However, we do not claim that agentic AI for hardware design is solved: these benchmarks are controlled proxies for a much broader engineering problem in chip design. Section~\ref{sec:discuss} examines the limitations of the current study and highlights open research challenges.
\end{abstract}

\section{Introduction}

Executable design tasks expose a limitation of single-turn code generation. A useful agent must place candidate artifacts inside a runnable workspace, invoke domain tools, interpret failures, and revise the artifacts until an explicit acceptance condition is satisfied. RTL design is a sharp test case: correctness depends on cycle-level behavior, reset and interface conventions, bit widths, and simulator feedback, so plausible Verilog is not enough.

Our motivation comes from repository-scale code self-evolution. AlphaEvolve showed that LLMs with automated evaluators can improve algorithmic kernels \citep{alphaevolve}; SATLUTION scaled the idea to full SAT-solver repositories \citep{satlution}; and ABCEvo applied it to the \textsc{ABC} logic-synthesis system \citep{abcevo}. In all cases, the agent evolves a version-controlled software artifact and admits changes only when executable evidence supports them. The missing step is hardware: prior repository-level self-evolution changes the programs that engineers run, not the hardware designs engineers create.

Here, we ask \textit{``whether hardware design itself can be managed as repository-level code evolution``}. Inspired by prior work \citep{abcevo,satlution}, \ours{} turns a design problem into a self-contained git worktree with an executable acceptance gate. A structured Markdown harness specifies the objective, domain knowledge, evaluator, and acceptance predicate; a bootstrap agent compiles it into a project pack; and a hands-free agent loop edits, evaluates, commits, or rejects candidate versions. Git is not incidental bookkeeping in this design. It provides the isolated evolving environment and the trace substrate: diffs expose state changes, commits define accepted checkpoints, logs and notes store evaluator evidence, and the repository history becomes a replayable record of the agent's search.

This framing is broader than the benchmarks in this paper. Agentic AI for hardware design includes architecture exploration, microarchitecture, verification planning, physical-design interaction, EDA software, and methodology development; we do not claim that RTL agents are solved. We use RTL benchmarks as controlled, executable proxies for studying whether repository-managed feedback can drive convergence. The evaluation spans ChipBench, RTLLM, Verilog-Eval, and nine CVDP categories, including completion, modification, reuse, testbench stimulus, checker and assertion generation, and debugging \citep{cvdp}.

This paper makes three contributions. First, we introduce \ours{}, a framework that hosts hardware design tasks as isolated, version-controlled, automatically evaluated repositories rather than as one-shot prompts. Second, we show that this git-native self-evolution loop can sweep complete RTL benchmark suites to a 100\% pass rate, with the only residual failure traced to a known specification-harness mismatch. Third, we analyze the resulting traces, including token consumption and test-generation coverage, and show that once executable feedback makes correctness converge, the main research bottleneck becomes convergence efficiency and verification quality.  

\section{Background and Related Work}

\paragraph{Why RTL is hard for language models.}
RTL generation differs from ordinary code completion because the output defines hardware that must satisfy temporal and bit-accurate behavior. A model must infer datapath widths, finite-state-machine transitions, reset conventions, ready-valid protocols, memory behavior, and corner cases that are often underspecified in natural language. A syntactically valid module is only a starting point; useful automation must connect generation to compilation, simulation, waveform or trace inspection, and repair.

\paragraph{RTL-specialized models and data.}
One body of work improves the generator itself. Early studies fine-tuned open models on Verilog corpora and established that domain adaptation matters: VeriGen curated large HDL training data and benchmarked open and closed models \citep{verigen}, RTLCoder released an open dataset and a lightweight model that surpassed GPT-3.5 on RTL generation \citep{rtlcoder}, and ChipNeMo domain-adapted LLMs across chip-design tasks \citep{chipnemo}. More recent efforts target data quality and reasoning: OriGen uses code-to-code augmentation with self-reflection \citep{origen}, CraftRTL constructs correct-by-construction synthetic data including non-textual representations \citep{craftrtl}, and ScaleRTL scales RTL reasoning data and adds test-time reasoning, improving Verilog-Eval and RTLLM \citep{scalertl}. These approaches strengthen first-attempt accuracy but do not, by themselves, define how an agent should iterate against an executable harness.

\paragraph{Iterative and agentic RTL.}
A complementary body of work adds tool use and iteration on top of a generator. AutoChip drives a generate-compile-simulate feedback loop \citep{autochip}; RTLFixer repairs syntax errors with retrieval-augmented, ReAct-style debugging \citep{rtlfixer}; VerilogCoder plans with a task-and-circuit-relation graph and traces waveforms via an AST-based tool to localize functional bugs \citep{verilogcoder}; MAGE decomposes a design across cooperating agents with high-temperature sampling and checkpoint-based debugging \citep{mage}; and ACE-RTL pairs an RTL-specialized generator with a frontier-model reflector and coordinator that evolves the prompting context over repair steps \citep{acertl}. These systems show that verification feedback substantially improves correctness, but each is built around a generation-and-repair pipeline for individual modules. \ours{} is complementary and more general: it is agnostic to the underlying generator and instead specifies how to \emph{host the entire problem as a versioned repository}, organize and gate the repair loop with native git operations, and drive a whole benchmark suite to completion, so any backbone can be evaluated on convergence rather than only on first-attempt accuracy.

\paragraph{Benchmarks for RTL design and verification.}
Verilog-Eval and RTLLM are widely used RTL generation benchmarks and remain useful for measuring basic specification-to-RTL ability \citep{verilogeval,rtllm}; ChipBench and consolidated suites such as OpenLLM-RTL aggregate further generation problems \citep{openllmrtl}. However, the CVDP paper argues that earlier suites are increasingly saturated and too narrow to represent real hardware design workflows \citep{cvdp}. CVDP contains 783 human-authored problems across 13 task categories. Its code-generation side includes RTL code completion, natural-language-specification to RTL, code modification, module reuse, linting or quality improvement, testbench stimulus generation, checker generation, assertion generation, and debugging. Its comprehension side includes RTL/specification correspondence, testbench/test-plan correspondence, and technical question answering.

CVDP also distinguishes non-agentic and agentic settings. Non-agentic problems provide the prompt and relevant context in a single turn, while agentic problems are packaged as mini-repositories intended for Dockerized agents that can inspect files and invoke tools. The benchmark reports 617 non-agentic and 166 agentic problems after quality filtering. The authors emphasize that state-of-the-art single-shot models struggle particularly on verification-oriented tasks such as testbench generation, checker generation, assertion generation, and bug fixing. This makes CVDP a strong fit for evaluating whether an agent can improve beyond first-attempt model accuracy through execution feedback.

\paragraph{Self-evolving agents over code repositories.}
A separate line of work treats the codebase itself as the object that an agent evolves. AlphaEvolve coupled an LLM with automated evaluators and an evolutionary loop to discover and refine algorithms at the scale of isolated kernels \citep{alphaevolve}. SATLUTION extended this to the full repository scale, evolving entire C/C++ SAT-solver repositories under strict correctness guarantees and distributed runtime feedback while also self-evolving its own evolution rules, ultimately outperforming the human-designed SAT-competition winners \citep{satlution}. ABCEvo carried the idea into EDA, using coordinated LLM agents to autonomously rewrite the million-line \textsc{ABC} logic-synthesis system under a correctness- and QoR-driven evaluation loop \citep{abcevo}. All three evolve EDA or scientific \emph{software}, the programs that engineers run, under the shared principle that a candidate change is useful only when it survives executable correctness checks and improves measured behavior. \ours{} carries the same repository-self-contained, evidence-gated principle to the \emph{hardware} side: rather than evolving a solver or synthesis kernel, it evolves the design under test, RTL sources, testbenches, and verification artifacts, as an automatically constructed task over a git worktree. This lets one protocol cover per-task RTL generation, completion, and repair, and is what allows \ours{} to treat an RTL benchmark problem ``as is'' rather than reformulating it into a software-engineering surrogate.

\paragraph{Git as an agent substrate.}
Closely related to our implementation is a recent trend of using version control itself as the scaffolding for LLM agents. EvoGit coordinates a population of decentralized coding agents purely through a Git-based phylogenetic graph that records the full version lineage, with no shared memory or explicit message passing \citep{evogit}. Git Context Controller elevates an agent's context from a transient token stream to a persistent, version-controlled memory with explicit \textsc{commit}, \textsc{branch}, and \textsc{merge} operations for long-horizon software tasks \citep{gcc}. Both demonstrate that git semantics, branching, lineage, and checkpointing, are a natural fit for organizing agentic exploration, but both target \emph{software} engineering: collaborative code generation and context management, respectively. \ours{} shares the conviction that git is the right substrate, and likewise records every attempt as a commit with attached evaluator evidence, but applies it to a different end, hosting an individual hardware-design problem as a self-contained, verification-gated repository whose history doubles as a replayable experience trace, which we view as convergent evidence that repository-native agents are an emerging paradigm rather than a single-domain trick.

\section{The HORIZON Framework}

\begin{figure}[H]
  \centering
  \makebox[\linewidth][c]{\includegraphics[width=1.12\linewidth]{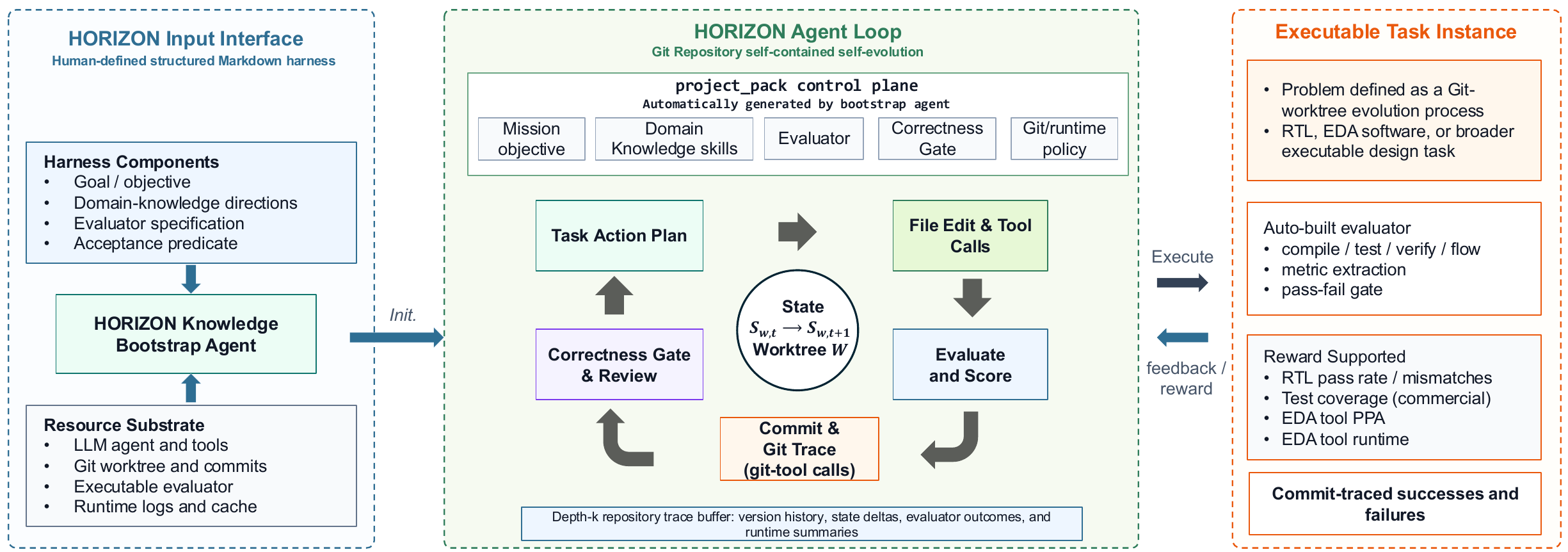}}
  \caption{Overview of \ours{}. A human-defined Markdown harness is converted into a project pack that specifies the mission, domain skills, evaluator, correctness gate, and git/runtime policy. The resulting task is solved by a self-contained agent loop over git-traced repository states. The executable task instance provides evaluator feedback and reward, while the depth-$k$ trace buffer records version history, state deltas, evaluator outcomes, and runtime summaries.}
  \label{fig:horizon-overview}
\end{figure}

\paragraph{System overview.}
Figure~\ref{fig:horizon-overview} summarizes \ours{}. The central idea is to manage a hardware-design problem like a software-evolution problem: the design, context, harness, and evaluator with correctness gate live in an isolated git worktree, and progress is expressed as repository state changes rather than as disconnected chat turns. A user provides a structured Markdown harness; a bootstrap agent compiles it into a \emph{project pack}, the control plane that fixes the mission, domain skills, executable evaluator, correctness gate, and git/runtime policy. From then on, a self-contained agent loop evolves the worktree without further human input. Each cycle generates or edits candidate artifacts, runs the evaluator, scores the result, and either commits the new version or logs the failure. Native git functions provide both isolation and traceability: diffs expose proposed state changes, commits define accepted checkpoints, logs recover the trajectory, and notes/runtime summaries attach evaluator evidence. The same machinery is intended to host versatile chip-design work, including RTL, EDA-software research, and methodology or flow exploration. In this paper we exercise the RTL instantiation; the broader design-space claim is a framework goal rather than a completed empirical validation.

\paragraph{From harness to executable task.}
\ours{} views language-model agents as policies acting on executable workspaces. The framework is not specific to RTL or to EDA self-evolution: any task with a persistent git worktree, machine-checkable feedback, and versioned artifacts can be organized in the same way. The only required user input is a structured Markdown harness, which may contain high-level intent, repository context, expected artifacts, evaluation criteria, and domain knowledge. Domain-aware harnesses are especially useful because they expose invariants, tool conventions, and failure modes that are difficult to infer from files alone.

The bootstrap agent converts this harness into a project pack. Let $m$ denote the input harness. A bootstrap tool loop $G_\phi$ constructs
\begin{equation}
  p = G_\phi(m), \qquad
  p = (\pi_{\mathrm{agent}}, E_p, A_p, \Gamma_p, \Omega_p),
\end{equation}
where $\pi_{\mathrm{agent}}$ is the agent policy prompt and tool contract, $E_p$ is an executable evaluator or harness, $A_p$ is the acceptance predicate, $\Gamma_p$ is the version-control and artifact policy, and $\Omega_p$ contains domain skills and repository instructions. For RTL, $E_p$ may include compilation, simulation, coverage extraction, and assertion or testbench checks. In other domains, the same slot may be filled by unit tests, theorem provers, profilers, security scanners, synthesis tools, or human-review gates. Problems are therefore defined over git worktrees rather than over a fixed target repository type.

\paragraph{Repository-traced formulation.}
\ours{} is an agentic system: the underlying policy is a free-form, history-dependent LLM agent, and we make no claim that its behavior is Markovian. We borrow the vocabulary of a semi-Markov decision process for one narrow purpose, to give precise, replayable names to the objects we record, not as a behavioral or optimization assumption. Because each accepted version follows a temporally extended episode of many edits, tool calls, and partial repairs, it is natural to label the boundaries: a ``state'' is a versioned snapshot of the repository (a bookkeeping checkpoint, not a sufficient statistic of the agent's reasoning), and an ``option'' is one such episode between two checkpoints. With that caveat, the objects below are simply definitions of what each trace stores. At outer checkpoint $t$, the recorded state is
\begin{equation}
  s_t = \big(\mathrm{tree}(w_t),\, p,\, z_t,\, \ell_{\le t},\, \mu_t\big),
\end{equation}
where $\mathrm{tree}(w_t)$ is the git tree of the current worktree, $p$ is the project pack, $z_t$ is campaign state, $\ell_{\le t}$ are accumulated logs and evaluator artifacts, and $\mu_t$ is any declared memory that the policy is allowed to condition on. The agent samples a variable-length option
\begin{equation}
  a_t = (\Delta_t,\, u_{t,1:K_t},\, \rho_t),
\end{equation}
where $\Delta_t$ is the proposed patch or generated artifact set, $u_{t,1:K_t}$ are the $K_t$ tool calls and observations inside the option, and $\rho_t$ is the final review or submission decision. The evaluator produces evidence
\begin{equation}
  y_t = E_p(w_t \oplus \Delta_t),
\end{equation}
and the acceptance predicate determines whether the trace advances:
\begin{equation}
  s_{t+1} =
  \begin{cases}
    \mathrm{Commit}(w_t \oplus \Delta_t,\, y_t,\, \Gamma_p), & A_p(y_t)=1, \\[2pt]
    \mathrm{RejectLog}(s_t,\, \Delta_t,\, y_t), & A_p(y_t)=0.
  \end{cases}
\end{equation}
The reward can be scalar or vector-valued, for example, 
\begin{equation}
  r_t = R_p(y_t) =
  \big[\Delta\mathrm{pass},\, \Delta\mathrm{coverage},\, \Delta\mathrm{QoR},\, -\mathrm{tokens},\, -\mathrm{time}\big],
\end{equation}
and an individual coordinate is populated only when the evaluator emits the corresponding signal; in this paper we report the $\Delta\mathrm{pass}$, $\Delta\mathrm{coverage}$, and $-\mathrm{tokens}$ components and leave synthesis quality-of-results ($\Delta\mathrm{QoR}$) to future work. An execution trace of depth $D$ is then
\begin{equation}
  \tau_{0:D} = \{(s_t,a_t,r_t,s_{t+1},y_t)\}_{t=0}^{D-1}.
\end{equation}
The depth $D$ is not fixed by the benchmark; it is determined by the campaign budget, convergence, or stopping rule. This makes the trace suitable for policy analysis, reward modeling, curriculum construction, or offline agent-RL training. We stress that we use this formulation only to structure and record the search; we do not train or update an RL policy in this work, and our agent backbone is fixed throughout a campaign.

\paragraph{Agent loop and trace buffer.}
Once the user supplies the initial Markdown harness, the loop is completely hands-free: bootstrap, generation, evaluation, acceptance, logging, and the next iteration all run automatically, and a campaign proceeds for many iterations with no further human intervention. Each outer transition contains an internal trajectory of depth $K_t$, the agent reads the current state, plans a target, edits the worktree, invokes tools, interprets failures, and either repairs or submits, and this inner trajectory is not assumed to be Markov and can differ in length at every step. We deliberately build the trace buffer on top of native git so that tracing is dynamic and essentially free to maintain: staged edits are inspected with \texttt{git diff --cached}, each accepted attempt becomes a \texttt{git commit} whose message and attached \texttt{git notes} carry the evaluator verdict and reward, the full version history is recovered with \texttt{git log}, and an independent review step diffs the candidate before it is allowed to submit. Successful commits become positive examples of repair strategies while rejected attempts are logged as negative examples of edits or tool-use paths that failed the evaluator, so the repository's own history \emph{is} the experience buffer rather than a separate datastore.

\paragraph{Memory and session reuse.}
Because there is no true Markov state, the process is just a sequence of agent actions and LLM responses, memory is handled pragmatically rather than as a state variable, and the operative objective is to maximize the share of \emph{cached} tokens relative to freshly billed input and output tokens. \ours{} reuses a persistent model session across iterations so that the harness, project pack, stable sources, and accumulated debugging context are served from the provider's prompt cache instead of being re-sent every turn; the newly billed tokens are then dominated by the current diff, the latest evaluator output, and the agent's response. This keeps the marginal cost of an additional repair iteration low even when a campaign runs for dozens of iterations, and it is the main reason cumulative token usage is overwhelmingly cached input (Section~\ref{sec:tokens}). The agent may still condition on session memory $h_t$,
\begin{equation}
  a_t \sim \pi_\theta(\cdot \mid s_t, h_t), \qquad
  h_{t+1}=M(h_t,s_t,a_t,y_t),
\end{equation}
but the source of truth remains the git worktree, the project pack, the evaluator outputs, and the versioned trace; campaign and review memories are kept separate so that review remains an independent check.

\section{Experiments}

\paragraph{Setup and protocol.}
\textbf{Model:} we use GPT-5.3 as the agent backbone for all experiments, fixed throughout;  \textbf{Benchmarks:} ChipBench, RTLLM-2.0, and Verilog-Eval, together with all CVDP code- and verification-generation categories (CID 002 to 016) spanning completion, specification-to-RTL, modification, reuse, linting/QoR, and stimulus, checker, and assertion generation as well as debugging. \textbf{Host environment:} all campaigns run on an AMD EPYC 9334 32-Core processor with 512\,GB of RAM, with evaluators invoking each suite's native open-source and, where required, commercial-EDA flows. \textbf{Task construction:} for each problem the bootstrap agent builds a project pack whose evaluator wraps the suite's native harness (compilation, simulation, and where available coverage or assertion checks), with the acceptance predicate set to the harness pass condition. \textbf{Protocol:} an \emph{iteration} is one automated outer step in which the agent edits the worktree, runs the evaluator, and either commits a passing version or logs a rejection; we report best-so-far pass rate, the fraction of tasks for which a passing version has been committed by a given iteration, and define the \emph{earliest-best} iteration as the first iteration that attains a benchmark's maximum observed pass rate. The entire loop is hands-free, and all results presented in this paper are obtained in single-agent mode.

\begin{table}[t]
  \centering
  \scriptsize
  \setlength{\tabcolsep}{3.2pt}
  \begin{tabular}{p{0.17\linewidth}p{0.28\linewidth}crr r}
    \toprule
    Suite/category & Evaluation focus & EDA & Iter. 0$^{b}$\, & Final iter. & \ours{} \\
    \midrule
    ChipBench & Mixed RTL generation tasks & Open & 20.0 & 5 & 100.0$^{a}$ \\
    RTLLM-2.0 & Natural-language spec to RTL & Open & 78.0 & 2 & 100.0 \\
    Verilog-Evalv2 & HDLBits-style Verilog generation & Open & 86.2 & 2 & 100.0 \\
    \midrule
    CVDP CID 002 & RTL code completion & Open & 3.2 & 82 & 100.0 \\
    CVDP CID 003 & Natural-language spec to RTL & Open & 19.2 & 24 & 100.0 \\
    CVDP CID 004 & RTL code modification & Open & 10.9 & 36 & 100.0 \\
    CVDP CID 005 & Spec-to-RTL module reuse & Open & 9.1 & 14 & 100.0 \\
    CVDP CID 007 & Linting / QoR improvement & Open & 0.0 & 24 & 100.0 \\
    CVDP CID 012 & Test-plan to stimulus generation & Comm. & 47.8 & 32 & 100.0 \\
    CVDP CID 013 & Test-plan to checker generation & Comm. & 3.8 & 19 & 100.0 \\
    CVDP CID 014 & Test-plan to assertion generation & Comm. & 79.1 & 1 & 100.0 \\
    CVDP CID 016 & Debugging and bug fixing & Open & 25.7 & 13 & 100.0 \\
    \midrule
    Overall & All evaluated RTL benchmarks &  & 47.8 &  & 100.0 \\
    \bottomrule
  \end{tabular}
\caption{Pass rates (\%) from a single \ours{} run.  \emph{Final iter.} denotes the iteration at which \ours{} converges. \emph{EDA} indicates the evaluation backend, open-source (Open) or commercial (Comm.); only CID~012--014 require a commercial simulator. \ours{} achieves 100\% completion on every suite. $^{a}$\,One non-passing ChipBench task is due to a specification--harness defect in the original benchmark; counting it as resolved yields 100\%. $^{b}$\,\emph{Iter.~0} is the pass rate after the first agent iteration, not the standalone LLM Pass@1.}
  \label{tab:benchmark-overview}

\end{table}

Table~\ref{tab:benchmark-overview} starts from the benchmark surface rather than only the final score. The evaluated tasks span compact RTL generation suites, legacy specification-to-RTL benchmarks, and nine CVDP categories that exercise completion, specification implementation, modification, module reuse, code improvement, testbench stimulus generation, checker generation, assertion generation, and debugging. 

\subsection{Benchmark completion and pass-rate progression}

Run as a single hands-free agentic loop per benchmark set, \ours{} drives every benchmark suite to a 100\% pass rate (Table~\ref{tab:benchmark-overview}); the only residual miss is a single ChipBench task, which we trace to a specification--harness mismatch in the original benchmark rather than to agent failure. What varies across suites is therefore not the destination but the path to it. At the agent's first iteration, the aggregate pass rate is 47.8\%, and it is substantially lower on the hardest CVDP categories, including 3.2\% on code completion (CID~002) and 3.8\% on checker generation (CID~013), before the same loop eventually closes the gap to 100\%. We emphasize that the iteration-0 results are not standalone model Pass@1 measurements. Instead, they reflect the state of the repository after the first step of the agentic evolution process, executed under the same prompting strategy and workflow used throughout the run. As a result, the agent may defer substantial exploration, debugging, and repair to later iterations rather than attempting to maximize first-pass success. This first-iteration aggregate is buoyed by Verilog-Eval-v2, which already reaches 86.2\% at iteration~0, whereas the CVDP subset starts at 23.9\%. We therefore report not merely that the suites are completed, but also how the agentic loop reaches completion.

\begin{figure}[t]
  \centering
  \begin{subfigure}[t]{0.49\linewidth}
    \centering
    \includegraphics[width=\linewidth]{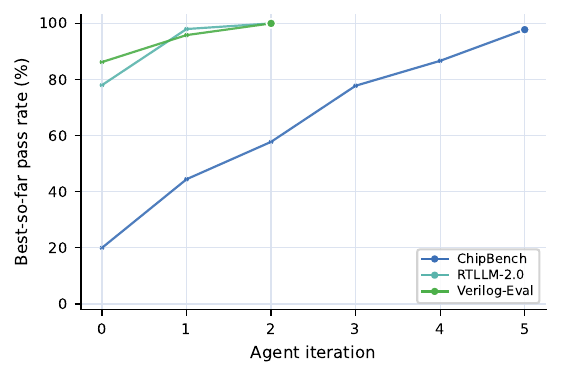}
    \caption{Simple RTL generation suites.}
  \end{subfigure}\hfill
  \begin{subfigure}[t]{0.49\linewidth}
    \centering
    \includegraphics[width=\linewidth]{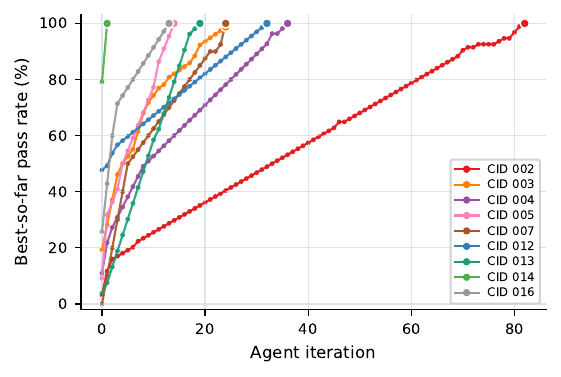}
    \caption{CVDP categories.}
  \end{subfigure}
  \caption{Best-so-far pass-rate trajectories over agent iterations. Earlier RTL generation suites saturate within a few iterations, while CVDP exposes longer repair trajectories and clearer differences in convergence difficulty.}
  \label{fig:pass-rate}
\end{figure}

Figure~\ref{fig:pass-rate} shows the qualitative difference between the benchmark families. RTLLM-2.0 and Verilog-Eval reach 100\% within two iterations; ChipBench climbs from 20.0\% to 100\% over five iterations, where the single task not passed under the original harness is the benchmark defect noted in Table~\ref{tab:benchmark-overview}. CVDP categories require more varied repair budgets. CID 014 reaches 100\% after one iteration, CID 016 and CID 005 reach 100\% in 13 and 14 iterations, CID 013 requires 19 iterations, CID 003 and CID 007 require 24 iterations, CID 012 requires 32 iterations, CID 004 requires 36 iterations, and CID 002 requires 82 iterations. The long tail in CID 002 is especially informative: it is not a one-shot modeling failure, but a convergence problem where the agent gradually converts many failing completions into passing designs.

The two extremes of difficulty are also the two most informative trajectories. CID 013 (adding reference-model checker logic to a testbench, evaluated under commercial-EDA simulation) rounds out the verification-generation family alongside stimulus generation (CID 012) and assertion generation (CID 014), and has the lowest first-iteration pass rate of any category, 3.8\%, consistent with the CVDP finding that checker writing is especially hard for single-shot models. Yet despite this weak start it reaches 100\% by iteration~19 along a strikingly steady, near-linear trajectory, climbing at a near-constant rate with almost no plateau. CID 013 and CID 002 thus bracket the behavior of the loop: a very low first-iteration rate does not by itself imply slow or unstable convergence (CID 013), while a long tail on a few stubborn designs is what actually drives cost (CID 002).

\subsection{Token Consumption Result}
\label{sec:tokens}

We next report how much agent work each completion requires, measured as the cumulative tokens consumed through a run's earliest-best iteration. This is not a normalized economic cost, model pricing, parallelism, and infrastructure are excluded, but it is a useful first-order measure of effort, and (per the session-reuse design) it is overwhelmingly cached input rather than freshly billed tokens.

\begin{table}[t]
  \centering
  \begin{minipage}[c]{0.50\linewidth}
    \centering
    \scriptsize
    \setlength{\tabcolsep}{4pt}
    \begin{tabular}{lrrr}
      \toprule
      Suite/category & Iter. & Tokens (M) & Share \\
      \midrule
      ChipBench & 5 & 2.8 & 1.3\% \\
      RTLLM-2.0 & 2 & 1.3 & 0.6\% \\
      Verilog-Evalv2 & 2 & 2.0 & 1.0\% \\
      CVDP CID 002 & 82 & 56.0 & 26.7\% \\
      CVDP CID 003 & 24 & 38.0 & 18.1\% \\
      CVDP CID 004 & 36 & 23.7 & 11.3\% \\
      CVDP CID 005 & 14 & 9.1 & 4.4\% \\
      CVDP CID 007 & 24 & 21.6 & 10.3\% \\
      CVDP CID 012 & 32 & 32.2 & 15.3\% \\
      CVDP CID 013 & 19 & 14.2 & 6.7\% \\
      CVDP CID 014 & 1 & 0.3 & 0.1\% \\
      CVDP CID 016 & 13 & 8.8 & 4.2\% \\
      \midrule
      Total &  & 209.9 & 100.0\% \\
      \bottomrule
    \end{tabular}
  \end{minipage}\hfill
  \begin{minipage}[c]{0.47\linewidth}
    \centering
    \includegraphics[width=\linewidth]{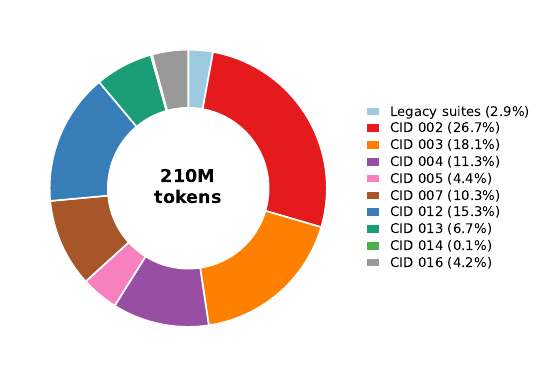}
  \end{minipage}
  \caption{Token consumption through the earliest-best iteration, as cumulative tokens recorded in the agent launch logs. Left: tokens (millions) and share of the total per benchmark, with the convergence iteration. Right: the same shares as a donut, with the three legacy suites grouped. Cost is dominated by a few hard CVDP categories. \textbf{Note that approximately 91\% of all tokens are cached input, which significantly lowered the API cost.} Shares may not sum to 100.0\% due to rounding.}
  \label{tab:token-usage}
\end{table}

\begin{figure}[t]
  \centering
  \begin{subfigure}[t]{0.49\linewidth}
    \centering
    \includegraphics[width=\linewidth]{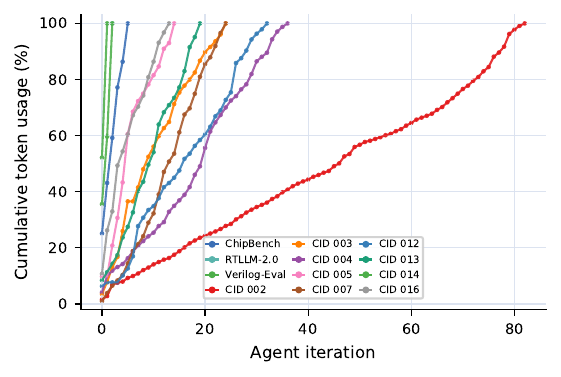}
    \caption{Normalized cumulative token usage.}
  \end{subfigure}\hfill
  \begin{subfigure}[t]{0.49\linewidth}
    \centering
    \includegraphics[width=\linewidth]{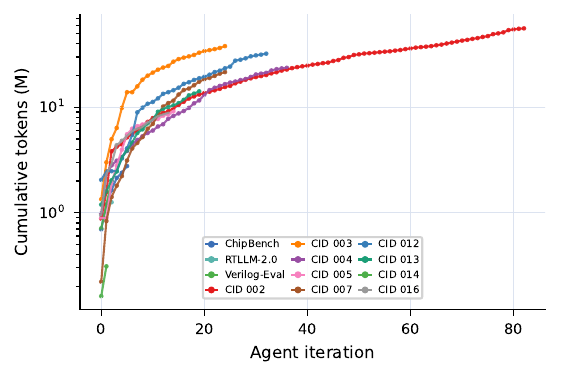}
    \caption{Absolute cumulative token usage.}
  \end{subfigure}
  \caption{Cumulative token-usage trajectories, truncated at each suite's convergence point (no spending beyond the best iteration is shown). The normalized view reports cumulative tokens as a percentage of the tokens consumed at convergence, while the absolute view reports cumulative tokens in millions on a log scale.}
  \label{fig:token-usage}
\end{figure}

Table~\ref{tab:token-usage} and Figure~\ref{fig:token-usage} show that token consumption is concentrated in the most challenging CVDP categories. The three legacy suites together consume 6.0M tokens, whereas the nine CVDP categories consume 203.9M tokens, accounting for 97.1\% of the total. Among these, CID 002 alone uses 56.0M tokens, CID 003 uses 38.0M, and CID 012 uses 32.2M. CID 013 is comparatively efficient given its difficulty: despite having the lowest first-iteration pass rate, it converges after consuming only 14.2M tokens, consistent with its steady, plateau-free trajectory.

The practical takeaway is that benchmark completion should be reported together with token consumption. Although \ours{} clears every suite, the final few percentage points on the most difficult categories absorb a disproportionate share of the budget. Consequently, we view token efficiency, rather than final pass rate, as the metric most in need of improvement. Notably, approximately 91\% of all tokens are cached input tokens. 

\subsection{Detailed discussion on test-generation tasks}

The test-generation categories deserve a closer look because they expose what \ours{} is and is not optimizing. For the two categories with parsed coverage data (CID 012 stimulus and CID 014 assertion generation), we additionally measure design coverage of the generated tests, reported as the average coverage percentage over designs with parsed coverage logs at each iteration. The crucial point is that \ours{}'s acceptance gate is the \emph{CVDP pass condition}, not a coverage target: the loop is driven to make the benchmark's own harness pass, and once a design passes, the gate is satisfied and the loop stops refining it. Coverage is therefore a secondary, observational signal here, it reports how much of the design the passing tests happen to exercise, rather than the objective being maximized.

\begin{table}[t]
  \centering
  \scriptsize
  \setlength{\tabcolsep}{4pt}
  \begin{tabular}{lrrrrr}
    \toprule
    Category & Iter. 0 pass$^{a}$\, &  Iter. 0 cov.$^{a}$\, & Best iter. & Best pass & Best cov. \\
    \midrule
    CVDP CID 012 & 47.8\% & 86.5\% & 32 & 100.0\% & 97.9\% \\
    CVDP CID 014 & 79.1\% & 98.1\% & 1 & 100.0\% & 100.0\% \\
    \bottomrule
  \end{tabular}
\caption{Coverage summary for the CVDP verification-generation categories. CID~012 is test-plan to testbench stimulus generation; CID~014 is test-plan to assertion generation. CID~014 has seven designs without parsed coverage rows at the best iteration, so its best-coverage average is computed over the remaining 60 parsed logs. $^{a}$\,Similarly as Table \ref{tab:benchmark-overview}, Iter.~0 denotes the first agent iteration and should not be interpreted as a standalone model Pass@1 or one-shot generation result; it reflects the repository state after the first step of the agentic evolution process.}

  \label{tab:coverage}
\end{table}

\begin{figure}[t]
  \centering
  \includegraphics[width=\linewidth]{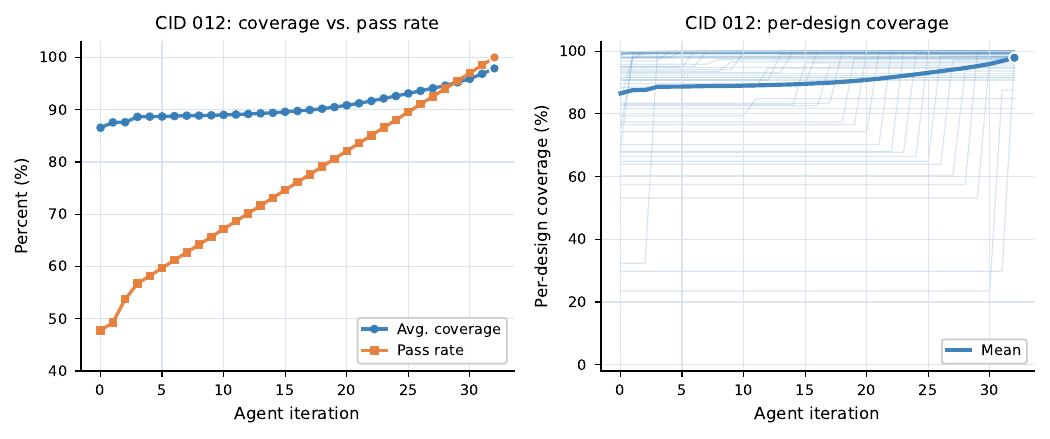}
  \caption{Coverage and pass rate for CID 012 (testbench stimulus generation), both truncated at iteration~32, where the pass rate first reaches 100\% (the convergence-point convention of Figures~\ref{fig:pass-rate}--\ref{fig:token-usage}). \textbf{Left}: average design coverage rises gradually from 86.5\% to 97.9\% while the pass rate climbs from 47.8\% to 100\% over the same iterations; the two move together but coverage plateaus below 100\% because the acceptance gate stops each design once it passes. \textbf{Right}: per-design coverage curves with their mean (bold). The improvement comes from lifting a low-coverage tail up toward full coverage, rather than only nudging already-high-coverage designs. 
  }
  \label{fig:coverage}
\end{figure}

Table~\ref{tab:coverage} and Figure~\ref{fig:coverage} make the stopping behavior concrete. CID 012 reaches a 100\% pass rate at iteration~32, but its average per-design coverage at that point is 97.9\%, not 100\%. This gap is expected and is a direct consequence of the acceptance gate: the loop halts on each design as soon as the CVDP harness passes, so coverage simply reflects the tests that were sufficient to pass rather than the maximum achievable. The per-design view (Figure~\ref{fig:coverage}, right) shows the loop lifting a low-coverage tail, several designs that begin near 20\% to 40\% coverage are pulled up toward full coverage, rather than only nudging already-high-coverage designs. CID 014 is the opposite regime: it starts near 98\% coverage and saturates immediately, so we report it only in Table~\ref{tab:coverage}; its 100.0\% best value is computed over the 60 designs with parsed coverage logs and should be read as coverage over those logs, not as evidence that every design emitted a usable report. We emphasize that we did \emph{not} attempt to drive coverage to 100\%: doing so would mean replacing the pass-based acceptance predicate with a coverage target, which \ours{} supports in principle but we leave to future work. Coverage here is thus a diagnostic that the generated tests are substantive, not a claim of exhaustive verification.

\section{Discussion and Limitations}
\label{sec:discuss}

The main takeaway of this work is that repository-managed executable feedback can make many RTL benchmark tasks converge. It is not that agentic hardware design is solved. The results should be read as \emph{benchmark convergence under the feedback made available to the agent}. Real chip projects involve incomplete specifications, changing constraints, downstream integration, human review, PPA tradeoffs, and validation targets that are not fully represented by current RTL benchmarks \cite{verilogeval,rtllm,openllmrtl,cvdp}.

The most important limitation is the reward-feedback interface. In the current \ours{} setup, the agent can inspect the outputs of each iterative evaluation. In CVDP \cite{cvdp}, for example, this includes simulator messages, evaluation logs, failure traces, and other task-local artifacts exposed by the benchmark harness. This mirrors a realistic debug workflow: engineers normally inspect logs and counterexamples, and rich feedback is what makes long-horizon repair feasible. At the same time, full access to these signals can create an \emph{over-solving} or reward-hacking failure mode. The agent may customize the generated RTL to match the observed failures, deterministic tests, or evaluator idiosyncrasies rather than implement the intended design semantics robustly. A passing result can therefore mean ``satisfies the visible harness under the exposed traces'' rather than ``satisfies the specification under all reasonable tests.'' This risk is especially relevant when benchmarks reveal detailed failure information or when the final acceptance test is the same harness used throughout the repair loop.

This raises a broader benchmarking issue. Existing RTL-agent benchmarks generally do not include a mechanism to detect over-solving or reward hacking \cite{verilogeval,rtllm,openllmrtl,cvdp}. They measure whether the submitted artifact passes the provided harness, but they usually do not separate debugging feedback from final hidden scoring, test robustness under randomized or perturbed stimuli, or audit whether a solution has specialized to artifacts of the evaluator. This is an open problem for the community. Future benchmarks for agentic hardware design should consider a two-level protocol: expose useful diagnostic feedback during repair, but reserve hidden randomized tests, independent reference models, formal equivalence checks, property suites, or held-out simulator configurations for final scoring. Reporting robustness to harness perturbations, coverage closure, and traces of what feedback the agent consumed would also make it easier to distinguish genuine design repair from benchmark-specific adaptation.

This tension has a direct parallel in software-engineering benchmarks. SWE-bench addresses it through structural test withholding \cite{jimenez2024swebench,wang2026solvedissues}. Agents receive a GitHub issue description and a repository snapshot at a fixed commit, while the fail-to-pass and pass-to-pass tests used for evaluation are withheld during inference and executed only after a final patch is submitted. This separation between repair-time information and evaluation-time scoring reduces opportunities for reward hacking and benchmark-specific adaptation. Subsequent analyses have shown that benchmark design choices such as solution leakage and weak test suites can substantially inflate measured performance \cite{aleithan2024swebenchplus}, and that patches deemed successful by benchmark tests may still diverge from developer-intended behavior or contain latent correctness issues \cite{swebench_experiments_issue16}. These findings suggest that future RTL-agent benchmarks should similarly separate diagnostic feedback from final scoring and incorporate robustness checks beyond the visible repair loop, such as hidden randomized tests, independent reference models, formal equivalence checking, or held-out verification environments.

Another major limitation is feedback turnaround. The RTL pass/fail benchmarks in this paper are relatively favorable because most evaluations complete quickly enough for iterative repair. In broader chip-design self-evolution, reward evaluation can be far slower. We have also studied PPA optimization in RTL design loops and PPA-oriented EDA-tool evolution, including the ABCEvo-style setting \cite{abcevo}, where the reward may require synthesis, placement, routing, timing analysis, power estimation, or large regression suites. SATLUTION already illustrates the cost of accurate repository-level reward: evaluating an entire SAT-competition benchmark required roughly a two-hour turnaround even with about 800 nodes running in parallel \cite{satlution}. For RTL PPA optimization or EDA-tool self-evolution, the turnaround can grow to days or weeks depending on design size, evaluation stage, and signoff fidelity. Long-latency reward fundamentally changes the agentic system problem: naive edit-evaluate-repair loops become too slow, and the agent must reason under delayed, sparse, and expensive feedback. Addressing long-turnaround reward is therefore a key research challenge for agentic chip design. 


\section{Conclusion}

We presented \ours{}, a self-evolving agent framework that treats hardware design as repository-level code evolution. A human-written Markdown harness is compiled into a project pack containing domain knowledge, an executable evaluator, an acceptance predicate, and a git/runtime policy; a hands-free agent loop then evolves an isolated repository worktree until the acceptance criterion is satisfied. Building on prior repository-scale self-evolution systems for EDA software, \ours{} extends the same paradigm to hardware-design artifacts themselves.

Across ChipBench, RTLLM, Verilog-Eval, and nine CVDP categories, \ours{} achieves 100\% benchmark completion with a fully hands-free agentic loop. To our knowledge, this is the first agentic system to complete all evaluated RTL benchmark suites end-to-end without human intervention. With correctness largely saturated on these benchmarks, the more informative signal becomes the cost of reaching that outcome. We find that token consumption is concentrated in a small number of difficult categories and that approximately 91\% of all tokens are cached input tokens, making token efficiency a more meaningful target for future improvement than final pass rate.

At the same time, we do not claim that agentic hardware design is solved. Current RTL benchmarks are controlled proxies for a much broader engineering problem and leave open important questions around reward hacking, robustness, long-horizon design planning, and deployment in production design flows. We hope this work helps establish a path from benchmark-scale RTL generation toward reliable agentic systems for real-world chip design.




{\small

\begin{thebibliography}{99}

\bibitem[Novikov et al.(2025)]{alphaevolve}
Alexander Novikov, Ng\^{a}n V\~{u}, Marvin Eisenberger, Emilien Dupont, Po-Sen Huang, et al.
\newblock AlphaEvolve: A Coding Agent for Scientific and Algorithmic Discovery.
\newblock arXiv preprint arXiv:2506.13131, 2025.

\bibitem[Yu et al.(2025)]{satlution}
Cunxi Yu, Rongjian Liang, Chia-Tung Ho, and Haoxing Ren.
\newblock Autonomous Code Evolution Meets NP-Completeness.
\newblock arXiv preprint arXiv:2509.07367, 2025.

\bibitem[Yu et al.(2026)]{abcevo}
Cunxi Yu, Rongjian Liang, Chia-Tung Ho, and Haoxing Ren.
\newblock Autonomous Evolution of EDA Tools: Multi-Agent Self-Evolved ABC.
\newblock arXiv preprint arXiv:2604.15082, 2026.

\bibitem[Pinckney et al.(2025)]{cvdp}
Nathaniel Pinckney, Chenhui Deng, Chia-Tung Ho, Yun-Da Tsai, Mingjie Liu, Wenfei Zhou, Brucek Khailany, and Haoxing Ren.
\newblock Comprehensive Verilog Design Problems: A Next-Generation Benchmark Dataset for Evaluating Large Language Models and Agents on RTL Design and Verification.
\newblock arXiv preprint arXiv:2506.14074, 2025.

\bibitem[Deng et al.(2026)]{acertl}
Chenhui Deng, Zhongzhi Yu, Guan-Ting Liu, Nathaniel Pinckney, Brucek Khailany, and Haoxing Ren.
\newblock ACE-RTL: When Agentic Context Evolution Meets RTL-Specialized LLMs.
\newblock arXiv preprint arXiv:2602.10218, 2026.

\bibitem[Deng et al.(2025)]{scalertl}
Chenhui Deng, Yun-Da Tsai, Guan-Ting Liu, Zhongzhi Yu, and Haoxing Ren.
\newblock ScaleRTL: Scaling LLMs with Reasoning Data and Test-Time Compute for Accurate RTL Code Generation.
\newblock arXiv preprint arXiv:2506.05566, 2025.

\bibitem[Liu et al.(2023)]{verilogeval}
Mingjie Liu, Nathaniel Pinckney, Brucek Khailany, and Haoxing Ren.
\newblock VerilogEval: Evaluating Large Language Models for Verilog Code Generation.
\newblock In \emph{Proceedings of the IEEE/ACM International Conference on Computer-Aided Design (ICCAD)}, 2023. arXiv:2309.07544.

\bibitem[Lu et al.(2024)]{rtllm}
Yao Lu, Shang Liu, Qijun Zhang, and Zhiyao Xie.
\newblock RTLLM: An Open-Source Benchmark for Design RTL Generation with Large Language Model.
\newblock In \emph{Proceedings of the Asia and South Pacific Design Automation Conference (ASP-DAC)}, 2024. arXiv:2308.05345.

\bibitem[Liu et al.(2025)]{openllmrtl}
Shang Liu, Yao Lu, Wenji Fang, Mengming Li, and Zhiyao Xie.
\newblock OpenLLM-RTL: Open Dataset and Benchmark for LLM-Aided Design RTL Generation.
\newblock In \emph{Proceedings of the IEEE/ACM International Conference on Computer-Aided Design (ICCAD)}, 2024. arXiv:2503.15112.

\bibitem[Thakur et al.(2024)]{verigen}
Shailja Thakur, Baleegh Ahmad, Hammond Pearce, Benjamin Tan, Brendan Dolan-Gavitt, Ramesh Karri, and Siddharth Garg.
\newblock VeriGen: A Large Language Model for Verilog Code Generation.
\newblock \emph{ACM Transactions on Design Automation of Electronic Systems}, 2024. arXiv:2308.00708.

\bibitem[Liu et al.(2025)]{rtlcoder}
Shang Liu, Wenji Fang, Yao Lu, Jing Wang, Qijun Zhang, Hongce Zhang, and Zhiyao Xie.
\newblock RTLCoder: Fully Open-Source and Efficient LLM-Assisted RTL Code Generation Technique.
\newblock \emph{IEEE Transactions on Computer-Aided Design of Integrated Circuits and Systems}, 2025. arXiv:2312.08617.

\bibitem[Liu et al.(2023)]{chipnemo}
Mingjie Liu, Teodor-Dumitru Ene, Robert Kirby, Chris Cheng, Nathaniel Pinckney, et al.
\newblock ChipNeMo: Domain-Adapted LLMs for Chip Design.
\newblock arXiv preprint arXiv:2311.00176, 2023.

\bibitem[Cui et al.(2024)]{origen}
Fan Cui, Chenyang Yin, Kexing Zhou, Youwei Xiao, Guangyu Sun, et al.
\newblock OriGen: Enhancing RTL Code Generation with Code-to-Code Augmentation and Self-Reflection.
\newblock arXiv preprint arXiv:2407.16237, 2024.

\bibitem[Liu et al.(2024)]{craftrtl}
Mingjie Liu, Yun-Da Tsai, Wenfei Zhou, and Haoxing Ren.
\newblock CraftRTL: High-quality Synthetic Data Generation for Verilog Code Models with Correct-by-Construction Non-Textual Representations and Targeted Code Repair.
\newblock arXiv preprint arXiv:2409.12993, 2024.

\bibitem[Thakur et al.(2023)]{autochip}
Shailja Thakur, Jason Blocklove, Hammond Pearce, Benjamin Tan, Siddharth Garg, and Ramesh Karri.
\newblock AutoChip: Automating HDL Generation Using LLM Feedback.
\newblock arXiv preprint arXiv:2311.04887, 2023.

\bibitem[Tsai et al.(2024)]{rtlfixer}
Yun-Da Tsai, Mingjie Liu, and Haoxing Ren.
\newblock RTLFixer: Automatically Fixing RTL Syntax Errors with Large Language Models.
\newblock In \emph{Proceedings of the 61st ACM/IEEE Design Automation Conference (DAC)}, 2024. arXiv:2311.16543.

\bibitem[Ho et al.(2025)]{verilogcoder}
Chia-Tung Ho, Haoxing Ren, and Brucek Khailany.
\newblock VerilogCoder: Autonomous Verilog Coding Agents with Graph-based Planning and Abstract Syntax Tree (AST)-based Waveform Tracing Tool.
\newblock In \emph{Proceedings of the AAAI Conference on Artificial Intelligence}, 2025. arXiv:2408.08927.

\bibitem[Zhao et al.(2025)]{mage}
Yujie Zhao, Hejia Zhang, Hanxian Huang, Zhongming Yu, and Jishen Zhao.
\newblock MAGE: A Multi-Agent Engine for Automated RTL Code Generation.
\newblock In \emph{Proceedings of the 62nd ACM/IEEE Design Automation Conference (DAC)}, 2025. arXiv:2412.07822.

\bibitem[Huang et al.(2025)]{evogit}
Beichen Huang, Ran Cheng, and Kay Chen Tan.
\newblock EvoGit: Decentralized Code Evolution via Git-Based Multi-Agent Collaboration.
\newblock arXiv preprint arXiv:2506.02049, 2025.

\bibitem[Wu et al.(2025)]{gcc}
Junde Wu, Jiayuan Zhu, and Yuyuan Liu.
\newblock Git Context Controller: Manage the Context of LLM-based Agents like Git.
\newblock arXiv preprint arXiv:2508.00031, 2025.

\bibitem[Jimenez et al.(2024)]{jimenez2024swebench}
Carlos E. Jimenez, John Yang, Alexander Wettig, Shunyu Yao, Kexin Pei, Ofir Press, and Karthik R. Narasimhan.
\newblock SWE-bench: Can Language Models Resolve Real-World GitHub Issues?
\newblock In \emph{Proceedings of the International Conference on Learning Representations (ICLR)}, 2024. arXiv:2310.06770.

\bibitem[Aleithan et al.(2024)]{aleithan2024swebenchplus}
Reem Aleithan, Haoran Xue, Mohammad Mahdi Mohajer, Elijah Nnorom, Gias Uddin, and Song Wang.
\newblock SWE-Bench+: Enhanced Coding Benchmark for LLMs.
\newblock arXiv preprint arXiv:2410.06992, 2024.

\bibitem[Wang et al.(2026)]{wang2026solvedissues}
You Wang, Michael Pradel, and Zhongxin Liu.
\newblock Are ``Solved Issues'' in SWE-bench Really Solved Correctly? An Empirical Study.
\newblock In \emph{Proceedings of the International Conference on Software Engineering (ICSE)}, 2026. Preprint available as arXiv:2503.15223.

\bibitem[pengfeigao1(2024)]{swebench_experiments_issue16}
pengfeigao1,
``Whether using test patch is allowed,''
\emph{SWE-bench/experiments}, GitHub issue \#16,
Jun. 7, 2024. Accessed: Jun. 23, 2026. [Online].
Available: \url{https://github.com/SWE-bench/experiments/issues/16}


\end{thebibliography}
}
\end{document}